\documentclass[12pt]{iopart}

\usepackage{graphicx}
\usepackage{amssymb,theorem}  
\eqnobysec
\newtheorem{theorem}{Theorem}

\newtheorem{lemma}{Lemma}
\begin{document}

\title[Analytically computable tangle for three-qubit mixed states]{Analytically computable tangle for three-qubit mixed states}

\author{Hiroyasu Tajima}

\address{Department of Physics, The University of Tokyo\\ 
4-6-1 Komaba, Meguro, Tokyo, 153-8505, Japan\\
TEL: +81-3-5452-6156 \quad FAX: +81-3-5452-6155\\
}
\ead{h-tajima@iis.u-tokyo.ac.jp}
\begin{abstract}
We present a new tripartite entanglement measure for three-qubit mixed states.
The new measure $t_{\mathrm{r}}(\rho)$, which we refer to as the r-tangle, is given as a kind of the tangle, but has a feature which the tangle does not have;
if we can derive an analytical form of $ t_{\mathrm{r}}(\rho)$ for a three-qubit mixed state $\rho$, we can also derive $t_{\mathrm{r}}(\rho')$ analytically for any states $\rho'$ which are SLOCC-equivalent to the state $\rho$. 
The concurrence of two-qubit states also satisfies the feature, but the tangle does not.
These facts imply that the r-tangle $t_{\mathrm{r}}$ is the appropriate three-partite counterpart of the concurrence.
We also derive an analytical form of the r-tangle $t_{\mathrm{r}}$ for mixtures of a generalized GHZ state and a generalized W state, and hence for all states which are SLOCC-equivalent to them.
\end{abstract}

\section{Introduction}
Quantum tasks beyond the classical tasks, such as quantum
computing, teleportation, superdense coding, $etc.$, utilize the entanglement as an important resource \cite{1,2,3,4}.
On one hand, with the development of the quantum information processing, manipulating many particles entangled to each other has become possible \cite{manipulate1,manipulate2}.
On the other hand, however, the quantification of the entanglement is still a fundamental problem in the field of quantum information.
Vigorous effort has been made, and the problem has been solved for two-qubit pure and mixed states as well as for three-qubit pure states. 
The concurrence \cite{5,6,7} and the negativity \cite{8} make it possible for us to quantify the entanglement analytically for two-qubit pure and mixed states. 
The stochastic LOCC classification of three-qubit pure states revealed \cite{GHZW} that there exist two types of three-partite entanglement, namely the GHZ-type and the W-type. The tangle \cite{10} and $J_{5}$ \cite{18} enabled us to quantify the entanglements of these two types.
With using the concurrence, the tangle, the parameter $J_{5}$ and the parameter $Q_{\mathrm{e}}$ introduced in Ref. \cite{tajima}, a necessary and sufficient condition of the possibility of deterministic LOCC transformations is given for arbitrary three-qubit pure states \cite{tajima}.

We thereby understood the features of two-qubit pure and mixed states as well as three-qubit pure states.
Apart from the above, however, our comprehension is not enough.
Although there have been many researches on the tangle for three-qubit mixed states,  its analytical form has been derived only in restricted regions \cite{t1,t2,Siebert,t3,t4}.
The approach for deterministic LOCC used in Ref. \cite{tajima} cannot be applied to three-qubit mixed states directly, because an important feature which holds for the tangle of pure states does $not$ hold for the tangle of mixed states; when we perform a measurement $\{M_{(i)}\}$ on the qubit $A$ of a three-qubit pure state $\left|\psi\right\rangle$, the tangle $\tau$ of the $i$th result $\left|\psi^{(i)}\right\rangle\equiv M_{(i)}\left|\psi\right\rangle/\sqrt{p_{(i)}}$ with the probability $p_{(i)}$ and the tangle $\tau$ of $\left|\psi\right\rangle$ satisfy the following equation:
\begin{eqnarray}
\tau(\left|\psi^{(i)}\right\rangle)=\alpha^2_{(i)}\tau(\left|\psi\right\rangle),\enskip\alpha_{(i)}\equiv\frac{\sqrt{\mathrm{det}M^\dagger_{(i)}M_{(i)}}}{p_{(i)}}.\label{multialphaprepre}
\end{eqnarray}  
This feature does not generally hold for the tangle of mixed states; we give an example that $\tau(\rho_{(i)})\ne\alpha^2_{(i)}\tau(\rho)$ in Appendix A.  

In the present paper, we introduce a new tripartite entanglement measure for three-qubit mixed states, which we refer to as the r-tangle.
The r-tangle $t_{\mathrm{r}}$ can be interpreted as a kind of the tangle; when the state is pure, the square of the r-tangle is equal to the tangle.
The r-tangle also satisfies the following equation: 
\begin{equation}
t_{\mathrm{r}}(\rho_{(i)})=\alpha_{(i)} t_{\mathrm{r}}(\rho),\enskip\rho_{(i)}\equiv \frac{M_{(i)}\rho M^\dag_{(i)}}{p_{(i)}}, \label{multialphapre}
\end{equation}
where $\alpha_{(i)}$ is the same as in \eref{multialphaprepre}.
The feature \eref{multialphapre} has two merits. 
First, using the r-tangle, we may be able to derive a necessary and sufficient condition of the possibility of deterministic LOCC transformations for arbitrary three-qubit mixed states;
because $t^2_{\mathrm{r}}(\rho_{(i)})=\alpha^2_{(i)} t^2_{\mathrm{r}}(\rho)$ holds, we may apply the approach in Ref. \cite{tajima} to the mixed states by employing $t^2_{\mathrm{r}}(\rho)$ as a substitute for the tangle $\tau(\rho)$.
Second, we can derive the r-tangle analytically in broader regions than the tangle;
if we can derive an analytical form of $t_{\mathrm{r}}(\rho)$ for a three-qubit mixed state $\rho$, the equation \eref{multialphapre} let us derive $t_{\mathrm{r}}(\rho')$ analytically for any states $\rho'$ which are SLOCC-equivalent to the state $\rho$.

Moreover, we also derive an analytical form of the r-tangle for mixtures of a generalized GHZ state and a generalized W state.
For such states, the analytical form of the tangle also has been derived \cite{Siebert}.
Using \eref{multialphapre}, we can derive the r-tangle not only for the mixtures but also for any states which are SLOCC-equivalent to the mixtures.
Note again that we also cannot apply the approach to the tangle, because $\tau(\rho_{(i)})=\alpha^2_{(i)}\tau(\rho)$ does not hold generally.

\section{Main Results}
In the present section, we give two theorems for the r-tangle for three-qubit mixed states.
First, we give the definition of the r-tangle:
\begin{equation}
t_{\mathrm{r}}(\rho)=\min_{\rho=\sum q_{i}\left|\psi_{i}\right\rangle\left\langle\psi_{i}\right|}\sum_{i}q_{i}\sqrt{\tau}(\left|\psi_{i}\right\rangle),\label{r-tangle}
\end{equation}
where $\sqrt{\tau}(\left|\psi\right\rangle)$ is written in terms of the coefficients $C_{ijk}$ as 
\begin{equation}
\left|\psi\right\rangle=\sum_{p,q,r}C_{pqr}\left|p q r\right\rangle
\end{equation}
\begin{eqnarray}
\sqrt{\tau}(\left|\psi\right\rangle)&=&\sqrt{4|d_{1}-2d_{2}+4d_{3}|},\\
d_{1}&=&C^2_{000}C^2_{111}+C^2_{001}C^2_{110}+C^2_{010}C^2_{101}+C^2_{100}C^2_{011},\\
d_{2}&=&C_{000}C_{111}C_{011}C_{100}+C_{000}C_{111}C_{101}C_{010}+C_{000}C_{111}C_{110}C_{001}\nonumber\\
&+&C_{011}C_{100}C_{101}C_{010}+C_{011}C_{100}C_{110}C_{001}+C_{101}C_{010}C_{110}C_{001},\\
d_{3}&=&C_{000}C_{110}C_{101}C_{011}+C_{111}C_{001}C_{010}C_{100}.
\end{eqnarray}

We refer to an ensemble $\{q_{i},\left|\psi_{i}\right\rangle\}$ of $\rho$ which minimizes the right-hand side of \eref{r-tangle} as the optimal ensemble.
We emphasize that $(t_{\mathrm{r}}(\rho))^2\ne\tau(\rho)$ because the mean of the square root is not equal to the square root of the mean. 
The equality $(t_{\mathrm{r}}(\rho))^2=\tau(\rho)$ is valid only when $\rho$ is pure.

Second, we give two theorems for the r-tangle.
The first theorem below means that when we obtain the value of the r-tangle and the optimal ensemble of a state $\rho$, then we also obtain them of any states which are S-LOCC equivalent to $\rho$. 
\begin{theorem}
Suppose that a measurement $\{M_{(j)}\}$ is performed on the qubit $A$ of an arbitrary three-qubit mixed state $\rho$ with the r-tangle $t_{\mathrm{r}}(\rho)$ and the optimal ensemble $\{q_{i},\left|\psi_{i}\right\rangle\}$.
Suppose also that the state $\rho_{(j)}=M_{(j)}\rho M^\dag_{(j)}/p_{(j)}$ is obtained as the $j$th result with the probability $p_{(j)}$.
The r-tangle $t_{\mathrm{r}}(\rho_{(j)})$ and the optimal ensemble $\{r_{i,(j)},\left|\varphi_{i,(j)}\right\rangle\}$ of $\rho_{(j)}$ are given by $t_{\mathrm{r}}(\rho)$ and $\{q_{i},\left|\psi_{i}\right\rangle\}$ as follows:
\begin{equation}
 t_{\mathrm{r}}(\rho_{(j)})=\alpha_{(j)} t_{\mathrm{r}}(\rho),\enskip\alpha_{(j)}\equiv\frac{\det\sqrt{M^\dag_{(j)} M_{(j)}}}{p_{(j)}}\label{multialpha},
\end{equation}
\begin{eqnarray}
r_{i,(j)}&=&\frac{q_{i}}{p_{(j)}}\left\langle\psi_{i}\right|M^\dag_{(j)} M_{(j)}\left|\psi_{i}\right\rangle,
\\
\left|\varphi_{i,(j)}\right\rangle&=&\frac{M_{(j)}\left|\psi_{i}\right\rangle}{\sqrt{\left\langle\psi_{i}\right|M^\dag_{(j)} M_{(j)}\left|\psi_{i}\right\rangle}}.
\end{eqnarray}
\end{theorem}

We can use Theorem 1 as follows;
when we obtain the r-tangle for a mixed state $\rho$, we can also obtain it for any states in the same SLOCC class as $\rho$.
Similarly, when we obtain the optimal ensemble for a mixed state $\rho$, we can also obtain it for any states in the same SLOCC class as $\rho$.
Theorem 1 does not hold for the tangle $\tau(\rho)$;
note again that $\tau(\rho)\ne( t_{\mathrm{r}}(\rho))^2$.
We show an explicit example of the case $\alpha^2_{(j)}\tau(\rho_{(j)})\ne\tau(\rho)$ in Appendix A. 

The second theorem below gives $ t_{\mathrm{r}}(\rho)$ analytically when $\rho$ is a mixture of generalized GHZ and generalized W states.
\begin{theorem}
We have
\begin{eqnarray}
 t_{\mathrm{r}}(\rho(p)) &=&\left\{ \begin{array}{ll}
0 & (0\le p\le p_{0}) \\
2|ab|\frac{p-p_{0}}{1-p_{0}} & (p_{0}\le p\le1) \\
\end{array} \right\},\label{abcdf}
\end{eqnarray}
\begin{eqnarray}
p_{0}&=&\frac{s^{2/3}}{1+s^{2/3}},\\
s&=&\frac{4cdf}{a^2b}>0,
\end{eqnarray}
for the family of three-qubit mixed states
\begin{equation}
\rho(p)=p\left|\mathrm{gGHZ}_{a,b}\right\rangle\left\langle\mathrm{gGHZ}_{a,b}\right|+(1-p)\left|\mathrm{gW}_{c,d,f}\right\rangle\left\langle\mathrm{gW}_{c,d,f}\right|,
\end{equation}
which consists of a generalized GHZ state
\begin{equation}
\left|\mathrm{gGHZ}_{a,b}\right\rangle=a\left|000\right\rangle+b\left|000\right\rangle,\enskip|a|^2+|b|^2=1
\end{equation}
and a generalized W state
\begin{equation}
\left|\mathrm{gW}_{c,d,f}\right\rangle=c\left|001\right\rangle+d\left|010\right\rangle+f\left|010\right\rangle,\enskip |c|^2+|d|^2+|f|^2=1.
\end{equation}
\end{theorem}

Note that the analytical form of $ t_{\mathrm{r}}(\rho(p))$ is simpler than that of $\tau(\rho(p))$ in Ref. \cite{Siebert}.
The r-tangle \eref{abcdf} consists of two straight lines as a function of $p$, whereas the function $\tau(\rho(p))$ in Ref. \cite{Siebert} consists of two straight lines and a curve.

\section{Proofs of Theorems}
\textbf{Proof of Theorem 1}:
We first consider the case in which $\rho$ is pure.
In this case, we have the equality $t_{\mathrm{r}}(\rho)=\sqrt{\tau(\rho)}$, and therefore Theorem 1 is included in Lemma 1 of Ref. \cite{tajima}.

Next, we consider the case in which $\rho$ is mixed.
Let us refer to the optimal ensembles of $\rho$ and $\rho_{(j)}$ as $\{q_{i},\left|\psi_{i}\right\rangle\}$ and $\{r_{k_{j}},\left|\varphi_{k_{j}}\right\rangle\}$, respectively.
We will prove that $t_{\mathrm{r}}(\rho_{(j)})=\alpha_{(j)}t_{\mathrm{r}}(\rho)$ and $\{r_{k_{j}},\left|\varphi_{k_{j}}\right\rangle\}=\{r_{i,(j)},\left|\varphi_{i,(j)}\right\rangle\}$.

First, we consider the case in which $\sqrt{\mbox{det}(M^{\dag}_{(j)}M_{(j)})}=0$ holds.
In the present case, the qubit $A$ becomes separable after the measurement, and thus the equation $ t_{\mathrm{r}}(\rho_{(j)})=0$ also holds. 
Thus, \eref{multialpha} is valid.
We can also prove that the ensemble $\{r_{i(j)},\left|\varphi_{i(j)}\right\rangle\}$ is optimal, because the states $\left|\varphi_{i(j)}\right\rangle$ are separable or biseparable states: the qubits $A$ of the states $\left|\varphi_{i(j)}\right\rangle$  are separable.
Thus Theorems 1 is valid when $\sqrt{\mbox{det}(M^{\dag}_{(j)}M_{(j)})}=0$ holds.
 
Second, let us consider the case in which $\sqrt{\mbox{det}(M^{\dag}_{(j)}M_{(j)})}\ne0$.
First, we show that if we can prove the following two equations, we can also prove Theorem 1:
\begin{eqnarray}
\alpha_{(j)}t_{\mathrm{r}}(\rho)\le\sum_{k_{j}}r_{k_{j}}t_{\mathrm{r}}(\left|\varphi_{k_{j}}\right\rangle),\label{1}\\
\sum_{i}r_{i(j)}t_{\mathrm{r}}(\left|\varphi_{i(j)}\right\rangle)\le\alpha_{(j)}\sum_{i}q_{i}t_{\mathrm{r}}(\left|\psi_{i}\right\rangle).\label{2}
\end{eqnarray}
Because $\{r_{k_{j}},\left|\varphi_{k_{j}}\right\rangle\}$ is the optimal ensemble of $\rho_{(j)}$ and because $\{r_{i,(j)},\left|\varphi_{i,(j)}\right\rangle\}$ is an ensemble of $\rho_{(j)}$, 
\begin{equation}
t_{\mathrm{r}}(\rho_{(j)})=\sum_{k_{j}}r_{k_{j}}t_{\mathrm{r}}(\left|\varphi_{k_{j}}\right\rangle)\le\sum_{i}r_{i(j)}t_{\mathrm{r}}(\left|\varphi_{i(j)}\right\rangle)
\end{equation}
is valid.
Note that $\{q_{i},\left|\psi_{i}\right\rangle\}$ is the optimal ensemble of $\rho$, and thus $t_{\mathrm{r}}(\rho)=\sum_{i}q_{i}t_{\mathrm{r}}(\left|\psi_{i}\right\rangle)$.
Thus, if \eref{1} and \eref{2} hold, 
\begin{equation}
\alpha_{(j)}t_{\mathrm{r}}(\rho)\le t_{\mathrm{r}}(\rho_{(j)})\le\sum_{i}r_{i(j)}t_{\mathrm{r}}(\left|\varphi_{i(j)}\right\rangle)\le\alpha_{(j)}t_{\mathrm{r}}(\rho)\label{cycle}
\end{equation}
also holds.
We can reduce \eref{cycle} to
\begin{equation}
\alpha_{(j)}t_{\mathrm{r}}(\rho)= t_{\mathrm{r}}(\rho_{(j)})=\sum_{i}r_{i(j)}t_{\mathrm{r}}(\left|\varphi_{i(j)}\right\rangle)=\alpha_{(j)}t_{\mathrm{r}}(\rho),
\end{equation}
and thus if we can prove Eqs. \eref{1} and \eref{2}, we can also prove Theorem 1.

Let us prove \eref{1} and \eref{2}.
First, we prove \eref{1}.
We prove \eref{1} by introducing an ensemble $\{r_{k_{j}}/L^2_{k_{j}},\left|\tilde{\varphi}_{k_{j}}\right\rangle\}$ of $\rho$ which satisfies
\begin{equation}
\alpha_{j}\sum_{k_{j}}\frac{r_{k_{j}}}{L^2_{k_{j}}}t_{\mathrm{r}}(\left|\tilde{\varphi}_{k_{j}}\right\rangle)=\sum_{k_{j}}r_{k_{j}}t_{\mathrm{r}}(\left|\tilde{\varphi}_{k_{j}}\right\rangle).\label{confor1}
\end{equation}
If we can introduce such ensemble of $\rho$, we can prove \eref{1} from \eref{confor1}; note that because $\{r_{k_{j}}/L^2_{k_{j}},\left|\tilde{\varphi}_{k_{j}}\right\rangle\}$ is an ensemble of $\rho$, 
\begin{equation}
t_{\mathrm{r}}(\rho)\le\sum_{k_{j}}\frac{r_{k_{j}}}{L^2_{k_{j}}}t_{\mathrm{r}}(\left|\tilde{\varphi}_{k_{j}}\right\rangle)
\end{equation}
is valid.

We obtain  $\{r_{k_{j}}/L^2_{k_{j}},\left|\tilde{\varphi}_{k_{j}}\right\rangle\}$ explicitly as follows.    
Now we consider the case in which $\mbox{det}(M_{(j)})\ne0$ holds, and thus we can take $M^{-1}_{(j)}$, which is the inverse of $M_{(j)}$.
We take the ensemble $\{r_{k_{j}}/L^2_{k_{j}},\left|\tilde{\varphi}_{k_{j}}\right\rangle\}$ as follows:
\begin{eqnarray}
\left|\tilde{\varphi}_{k_{j}}\right\rangle\equiv L_{k_{j}}\sqrt{p_{(j)}}M^{-1}_{(j)}\left|\varphi_{k_{j}}\right\rangle,
\end{eqnarray}
where $L_{k_{j}}$ are normalization constants.
We can prove that $\{r_{k_{j}}/L^2_{k_{j}},\left|\tilde{\varphi}_{k_{j}}\right\rangle\}$ satisfies \eref{confor1}, as follows;
\begin{eqnarray}
\sum_{k_{j}}r_{k_{j}} t_{\mathrm{r}}(\left|\varphi_{k_{j}}\right\rangle)
=\sum_{k_{j}}r_{k_{j}} t_{\mathrm{r}}\left(\frac{M_{(j)}}{L_{k_{j}}\sqrt{p_{(j)}}}\left|\tilde{\varphi}_{k_{j}}\right\rangle\right)\nonumber\\
=\sum_{k_{j}}r_{k_{j}}\frac{\sqrt{\mbox{det}(M^\dag_{j}M_{(j)})}}{L^2_{k_{j}}p_{(j)}} t_{\mathrm{r}}(\left|\tilde{\varphi}_{k_{j}}\right\rangle)=\alpha_{(j)}\sum_{k_{j}}\frac{r_{k_{j}}}{L^2_{k_{j}}} t_{\mathrm{r}}(\left|\tilde{\varphi}_{k_{j}}\right\rangle).
\end{eqnarray}

Finally, let us prove \eref{2}. 
Note that we can write $\{r_{i(j)},\left|\varphi_{i(j)}\right\rangle\}$ as 
\begin{equation}
r_{i(j)}=\frac{q_{i}}{N^2_{ij}},\enskip\left|\varphi_{i(j)}\right\rangle= N_{ij}\frac{M_{(j)}}{\sqrt{p_{(j)}}}\left|\psi_{i}\right\rangle,\enskip N_{ij}\equiv \frac{\sqrt{p_{(j)}}}{\sqrt{\left\langle\psi_{i}\right|M^\dag_{(j)} M_{(j)}\left|\psi_{i}\right\rangle}}.
\end{equation}
Thus, we can derive \eref{2} as follows:
\begin{eqnarray}\fl
\sum_{i}r_{i(j)} t_{\mathrm{r}}(\left|\varphi_{i(j)}\right\rangle)=\sum_{i}\frac{q_{i}}{N^2_{ij}} t_{\mathrm{r}}\left(N_{ij}\frac{M_{(j)}}{\sqrt{p_{(j)}}}\left|\psi_{i}\right\rangle\right)\nonumber\\\fl
=\sum_{i}\frac{q_{i}}{N^2_{ij}}\frac{N^2_{ij}\sqrt{\mbox{det}(M^\dag_{(j)}M_{(j)})}}{p_{(j)}} t_{\mathrm{r}}(\left|\psi_{i}\right\rangle)=\frac{\sqrt{\mbox{det}(M^\dag_{(j)}M_{(j)})}}{p_{(j)}}\sum_{i}q_{i} t_{\mathrm{r}}(\left|\psi_{i}\right\rangle).
\end{eqnarray}
This completes the proof of Theorem 1.
$\Box$

\textbf{Proof of Theorem 2}:
We prove the present theorem by a method similar to the one used in Ref. \cite{Siebert}.
First, we prove the following lemma.
\begin{lemma}
If there is a function $f(p)$ which satisfies the following three conditions, it must be $ t_{\mathrm{r}}(\rho(p))$:
\begin{description}
\item[Condition 1]{The following inequality holds for any $p$ and $\varphi$:
\begin{equation}
f(p)\le t_{\mathrm{r}}(\left|p,\varphi\right\rangle),\label{condition1}
\end{equation}
where
\begin{eqnarray}
\left|p,\varphi\right\rangle&\equiv&\sqrt{p}\left|gGHZ_{a,b}\right\rangle+\sqrt{1-p}e^{i(\varphi-\tilde{\varphi}/3)}\left|gW_{c,d,f}\right\rangle,\\
\tilde{\varphi}&\equiv& \mathrm{arg}\left[\frac{4cdf}{a^2b}\right].
\end{eqnarray}}
\item[Condition 2]{There exists an ensemble $\{p_{i},\left|q_{i},\varphi_{i}\right\rangle\}$ of $\rho(p)$ which satisfies the following equation: 
\begin{equation}
f(p)=\sum_{i}p_{i} t_{\mathrm{r}}(\left|q_{i},\varphi_{i}\right\rangle).\label{condition2}
\end{equation}}
\item[Condition 3]{The function $f(p)$ is a convex function.}
\end{description}
\end{lemma}
\textit{Proof}: Because of Condition 2, we have $f(p)\ge t_{\mathrm{r}}(\rho(p))$.
We also prove $f(p)\le t_{\mathrm{r}}(\rho(p))$ as follows:
\begin{eqnarray}
 t_{\mathrm{r}}(\rho(p))&=&\sum_{i}\tilde{p}_{i} t_{\mathrm{r}}(\left|\tilde{q}_{i},\tilde{\varphi}_{i}\right\rangle)\nonumber\\
&\ge&\sum_{i}\tilde{p}_{i}f(\tilde{q}_{i})\ge f(\sum_{i}\tilde{p}_{i}\tilde{q}_{i})=f(p),
\end{eqnarray}
where $\{\tilde{p}_{i},\left|\tilde{q}_{i},\tilde{\varphi}_{i}\right\rangle\}$ is the optimal ensemble of $\rho(p)$. We have derived the first inequality from Condition 1 and the second inequality from Condition 3. ($\Box$)

Now we only have to prove that the right-hand side of \eref{abcdf}, which we refer to as $g(p)$, satisfies Conditions 1--3.
First, the function $g(p)$ is clearly convex, and thus Condition 3 holds.
Second, we can take the ensemble of $\rho(p)$ which satisfies \eref{condition2} as follows:
\begin{eqnarray}\fl
\rho(p) =\left\{\begin{array}{ll}
\frac{p_{0}-p}{p_{0}}\left|0,0\right\rangle\left\langle 0,0\right|+\frac{p}{3p_{0}}\sum^{2}_{n=0}\left|p_{0},\frac{2n\pi}{3}\right\rangle\left\langle p_{0},\frac{2n\pi}{3}\right| & (0\le p\le p_{0}) \\
\frac{p-p_{0}}{1-p_{0}}\left|1,0\right\rangle\left\langle 1,0\right|+\frac{1-p}{3(1-p_{0})}\sum^{2}_{n=0}\left|p_{0},\frac{2n\pi}{3}\right\rangle\left\langle p_{0},\frac{2n\pi}{3}\right| & (p_{0}\le p\le1) \\
\end{array} \right\}.\label{ensemble}
\end{eqnarray}
To prove that the above ensembles satisfy \eref{condition2}, we only have to notice that
\begin{eqnarray}
 t_{\mathrm{r}}(\left|p,\varphi\right\rangle)=2|ab|\sqrt{\left|p^2-\sqrt{p(1-p)^3}e^{3i\varphi}\frac{4cdf}{a^2b}\right|},\label{amountoftau}
\end{eqnarray}
and especially
\begin{eqnarray}
 t_{\mathrm{r}}(\left|1,0\right\rangle)&=&2|ab|,\label{amountof10}\\
 t_{\mathrm{r}}(\left|0,0\right\rangle)&=& t_{\mathrm{r}}(\left|p_{0},\frac{2n\pi}{3}\right\rangle)=0.\label{amountof00}
\end{eqnarray}
Thus, Condition 2 is valid.

\begin{figure}
\begin{center}
\includegraphics[width=100mm]{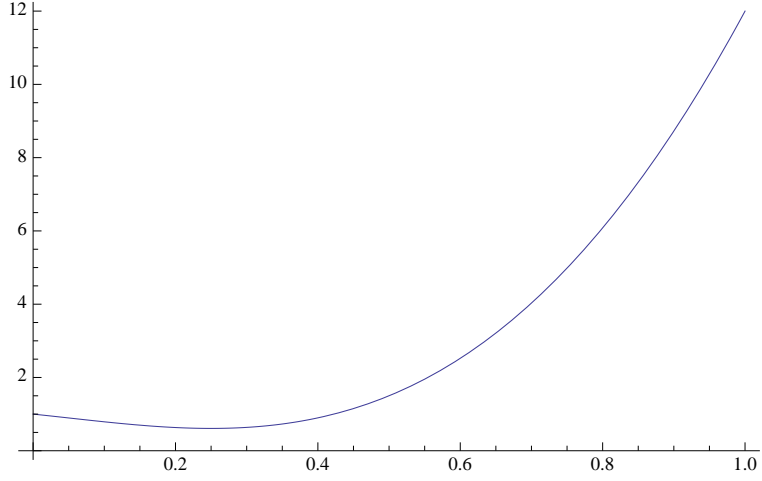}
\end{center}
\caption{The graph of the function $-4p^4+20p^3-3p^2-2p+1$ from 0 to 1.}
\label{kensyouzu}
\end{figure}

Finally, let us prove Condition 1.
Because $ t_{\mathrm{r}}$ is non-negative, $g(p)$ clearly satisfies Condition 1 for $0\le p\le p_{0}$.
Note that for $p_{0}\le p\le1$, the function $g(p)$ is a linear function of $p$ and that the following three expressions hold:
\begin{eqnarray}
 t_{\mathrm{r}}(\left|p,0\right\rangle)&\le& t_{\mathrm{r}}(\left|p,\varphi\right\rangle),\\
 t_{\mathrm{r}}(\left|1,0\right\rangle)&=&g(1),\\
 t_{\mathrm{r}}(\left|0,0\right\rangle)&=&g(p_{0}).
\end{eqnarray}
Thus, if we can prove $ t_{\mathrm{r}}(\left|p,0\right\rangle)$ is concave for $p_{0}\le p\le 1$, then we can also prove $g(p)\le t_{\mathrm{r}}(\left|p,\varphi\right\rangle)$ for $p_{0}\le p\le 1$.
Let us prove the concaveness of $ t_{\mathrm{r}}(\left|p,0\right\rangle)$.
Only for simplicity, we refer to $4cdf/(a^2b)$ as $s$.
Then, 
\begin{eqnarray}\fl
&&\frac{d^2}{dp^2} t_{\mathrm{r}}(\left|p,0\right\rangle)=2|ab|\frac{d^2}{dp^2}\sqrt{p^2-\sqrt{p(1-p)^3}s}\\\fl
&=&-\frac{1}{4( t_{\mathrm{r}}(\left|p,0\right\rangle))^3}\left(s\frac{12p^3+3p/2}{\sqrt{p(1-p)}}+s^2\frac{-4p^4+20p^3-3p^2-2p+1}{4p(1-p)}\right).\label{twoterms}
\end{eqnarray}
The term $12p^3+3p/2$ is clearly positive.
The term $-4p^4+20p^3-3p^2-2p+1$ is also positive for $0\le p\le1$ as shown in Fig\ref{kensyouzu}.
Therefore, $t_{\mathrm{r}}(\left|p,0\right\rangle)$ is concave for $p_{0}\le p\le 1$, and thus the function $g(p)$ satisfies Conditions 1--3.
Hence, because of Lemma 1, the equation $t_{\mathrm{r}}(\rho(p))=g(p)$ is valid.
$\Box$

\section{Conclusion}
In the present article, we introduced a new entangleemnt measure which we call the r-tangle.
The r-tangle $t_{\mathrm{r}}$ satisfies $ t_{\mathrm{r}}(\rho_{(j)})=\alpha_{(j)} t_{\mathrm{r}}(\rho)$.
Thanks to the feature, if we derive an analytical form of the r-tangle for a three-qubit mixed state, we can also derive the r-tangle analytically for any states which are SLOCC-equivalent to the state.
Note that the concurrences also satisfy a similar feature $C(\rho_{(j)})=\alpha_{(j)}C(\rho)$, and that the tangle $\tau(\rho)$ does $not$ satisfies such a feature; we show an example that $\tau(\rho_{(j)})\ne\alpha^2_{(j)}\tau(\rho)$ in Appendix A.
These facts imply that we should consider the r-tangle instead of the tangle as the three-partite counterpart of the concurrence.
Moreover, we derive the analytical form of the r-tangle for mixtures of generalized GHZ state and generalized W state.
Although the tangle has been also derived for such states \cite{Siebert}, the form of $t_{\mathrm{r}}(\rho(p))$ is simpler than that of  $\tau(\rho(p))$ as the function of $p$.
Using $ t_{\mathrm{r}}(\rho_{(j)})=\alpha_{(j)} t_{\mathrm{r}}(\rho)$, we can derive the r-tangle not only for the mixture but also for any state which is SLOCC-equivalent to the mixtures.
We cannot apply the approach to the tangle, because $\tau(\rho_{j})=\alpha^2_{(j)}\tau(\rho)$ does not hold generally.

\section*{Acknowledgements}
This work was supported by the Grants-in-Aid for Japan Society for Promotion of Science (JSPS) Fellows (Grant No. 24E8116).
The author thanks Prof. Naomichi Hatano for useful discussions.

\appendix

\section{}
In the present appendix, we will show a counterexample of $\tau(\rho_{(j)})=\alpha^2_{(j)}\tau(\rho)$.
Let us consider the following three-qubit mixed state:
\begin{equation}\fl
\rho=\frac{4}{5}\left|\mathrm{gGHZ}_{1/\sqrt{2},1/\sqrt{2}}\right\rangle\left\langle\mathrm{gGHZ}_{1/\sqrt{2},1/\sqrt{2}}\right|+\frac{1}{5}\left|\mathrm{gW}_{1/\sqrt{3},1/\sqrt{3},1/\sqrt{3}}\right\rangle\left\langle\mathrm{gW}_{1/\sqrt{3},1/\sqrt{3},1/\sqrt{3}}\right|.
\end{equation}
According to a result in Ref. \cite{Siebert}, we have $\tau(\rho)=(63-\sqrt{465})/90$.
Let us perform the following measurement on the qubit $A$ of $\rho$:
\begin{eqnarray}
M_{(0)}=\left(
\begin{array}{cc}
1 & 0  \\
 0 & \frac{1}{\sqrt{10}} \end{array}
\right),\enskip\enskip\enskip
M_{(1)}=\left(
\begin{array}{cc}
0 & 0  \\
0 & \frac{3}{\sqrt{10}} \end{array}
\right).
\end{eqnarray}
The probability $p_{(0)}$ that we obtain the result 0 is $29/50$, for which the state becomes
\begin{eqnarray}
\rho_{(0)}=\frac{22}{29}\left|\mathrm{gGHZ}_{\sqrt{10/11},\sqrt{1/11}}\right\rangle\left\langle\mathrm{gGHZ}_{\sqrt{10/11},\sqrt{1/11}}\right|\nonumber\\
+\frac{7}{29}\left|\mathrm{gW}_{\sqrt{10/21},\sqrt{10/21},\sqrt{1/21}}\right\rangle\left\langle\mathrm{gW}_{\sqrt{10/21},\sqrt{10/21},\sqrt{1/21}}\right|.
\end{eqnarray}
According to the result in Ref. \cite{Siebert}, $\tau(\rho_{(0)})=160(9-\sqrt{6})/7569$.
Thus,
\begin{eqnarray}
\alpha^2_{(0)}=\frac{\mbox{det}M^\dagger_{(0)} M_{(0)}}{p^2_{(0)}}=\frac{250}{841}\nonumber\\
\ne\frac{1600(9-\sqrt{6})}{841(63-\sqrt{465})}=\frac{\tau(\rho_{(0)})}{\tau(\rho)}.
\end{eqnarray}
$\Box$

\end{document}